\documentclass[aps,prb,twocolumn,showpacs,floatfix,superscriptaddress]{revtex4}
\usepackage{graphicx,bm,amsmath,amssymb,natbib,url,epsfig,psfrag}

\begin{document}
\title{Geometric phase at graphene edge}
\author{Sang-Jun Choi, Sunghun Park, H.-S. Sim}
\affiliation{Department of Physics, Korea Advanced Institute of
Science and Technology, Daejeon 305-701, Korea}

\date{\today}

\begin{abstract}
We study the scattering phase shift of Dirac fermions at graphene edge. We find that when a plane wave of a Dirac fermion is reflected at an edge of graphene, its reflection phase is shifted by the geometric phase resulting from the change of the pseudospin of the Dirac fermion in the reflection. The geometric phase is the Pancharatnam-Berry phase that equals the half of the solid angle on Bloch sphere determined by the propagation direction of the incident wave and also by the orientation angle of the graphene edge. The geometric phase is finite at zigzag edge in general, while it always vanishes at armchair edge because of intervalley mixing. To demonstrate its physical effects, we first connect the geometric phase with the energy band structure of graphene nanoribbon with zigzag edge. The magnitude of the band gap of the nanoribbon, that opens in the presence of the staggered sublattice potential induced by edge magnetization, is related to the geometric phase. Second, we numerically study the effect of the geometric phase on the Veselago lens formed in a graphene nanoribbon. The interference pattern of the lens is distinguished between armchair and zigzag nanoribbons, which is useful for detecting the geometric phase.
\end{abstract}



\maketitle


\section{Introduction}

Dirac fermions in graphene, that describe electronic properties in the low energy regime,~\cite{Wallace,Semenoff,DiVincenzo} have Berry phase,~\cite{Berry,Anandan} because of momentum-pseudospin locking~\cite{CastroNeto}; the pseudospin represents the sublattice sites of the unit cell of graphene. They acquire Berry phase $\pi$ when they propagate along a close trajectory. This topological effect results in unusual phenomena such as the half-integer quantum Hall effect,~\cite{Novoselov1,Zhang} Klein tunneling,~\cite{Katsnelson} and weak antilocalization.~\cite{Suzuura,Morozov,Tikhonenko}

Recently, a Berry-phase scattering effect of Dirac fermions was predicted by the authors of this paper.~\cite{Choi} This Berry phase occurs as a scattering phase shift in a single scattering event of transmission or reflection of Dirac fermions at a junction with spatially nonuniform mass gap. It can be tuned to an arbitrary value (i.e., not fixed to $\pi$) by junction control. It provides a tool of detecting the Chern number of a Dirac-fermion insulator with mass gap, and it contributes to the quantization rule of Dirac fermions, suggesting geometric-phase devices with nontrivial charge and spin transport. This geometric phase of Dirac fermions is an electronic analogue of Pancharatnam-Berry phase~\cite{Pancharatnam, BerryPan, Bhandari, Ben-Aryeh} of polarized light, which occurs when polarized light passes through a series of optical polarizers. In this work, we predict that this geometric phase also appears at graphene edge.

On the other hand, graphene nanoribbons have attracted attention. 
The properties of their electronic structure, including the energy band gap at Fermi level, depend on its width and edge orientation angle.~\cite{Fujita, Yamashiro, Son, Ezawa, Han, Yazyev, Akhmerov1} For instance, an armchair graphene ribbon is metallic or semiconducting, depending on the ribbon width, while zigzag ribbons are always metallic when electron interactions are negligible.~\cite{Fujita} In zigzag ribbons, Coulomb interactions cause the instability of the flat band of the edge states, and result in edge magnetization and band gap opening.~\cite{Yamashiro, Son} Moreover, it has been shown that the pristine zigzag edge is energetically metastable and its hexagonal lattice structure is reconstructed into pentagon-heptagon pairs, Stone-Wales defects,~\cite{Koskinen1, Koskinen2, Girit, Wassmann, Huang} which alter the edge state from a dispersionless band into a dispersive band.~\cite{Koskinen1} It will be interesting to see topological aspects of these edge effects in terms of the geometric phase of Dirac fermions.



In this paper, we predict the topological nature of the reflection of Dirac fermions at graphene edge. When a plane wave of a Dirac fermion is reflected at a graphene edge, the reflection phase shift is contributed by the geometric phase resulting from the change of the pseudospin of the Dirac fermion in the reflection. The geometric phase is the Pancharatnam-Berry phase that equals the half of the solid angle on Bloch sphere, which is determined by the propagation direction of the incident wave and the orientation angle of the graphene edge. The geometric phase is finite at zigzag edge in general, while it always vanishes at armchair edge because of intervalley mixing. 

To demonstrate the effects of the geometric phase, we first connect the geometric phase with the energy band structure of a zigzag graphene nanoribbon. The geometric phase provides the topological view why the electronic structure of a zigzag nanoribbon is very different from that of an armchair nanoribbon. The magnitude of the band gap of a zigzag nanoribbon, that opens in the presence of the staggered sublattice potential~\cite{Son, Akhmerov1} induced by edge magnetization, is related to the geometric phase. Second, we numerically study the effect of the geometric phase on a Veselago lens~\cite{Cheianov} formed in a nanoribbon, using a recursive Green function method.~\cite{Sim,Xing} Because of the geometric phase, the interference pattern (i.e., caustics pattern) of the lens in a zigzag nanoribbon is different from the case of a graphene sheet where an edge effect is ignorable. In contrast, the armchair case shows the identical feature to the graphene sheet. This behavior is useful for detecting the geometric phase.

This paper is organized as follows. In Sec.~\ref{sec:Model}, we briefly introduce the boundary condition of graphene edge. In Sec.~\ref{sec:PBphase}, we obtain the reflection phase shift and the Pancharatnam-Berry geometric phase at graphene edge. In Sec.~\ref{sec:ZGNR}, we connect the energy band gap of a zigzag graphene nanoribbon with the geometric phase. In Sec.~\ref{sec:Veselago}, we present numerical results for Veselago lens. The summary is given in Sec.~\ref{sec:Summary}.

\section{graphene edge}
\label{sec:Model}


We will predict Pancharatnam-Berry geometric phase of Dirac fermions at graphene edge, by using the low-energy continuum Hamiltonian and the boundary condition of Dirac fermions at graphene edge; we numerically confirm the effect of the geometric phase in Sec.~\ref{sec:Veselago}, based on the tight-binding lattice Hamiltonian of graphene. The boundary condition was obtained in Refs.~\cite{McCann, Akhmerov1,Brey,Ostaay}. In this section, we briefly introduce the boundary condition. 

In the low-energy regime, graphene is effectively described by the Dirac Hamiltonian~\cite{Wallace,Semenoff, DiVincenzo}.
In the valley isotropic representation~\cite{Akhmerov2}, it is written as
\begin{equation}\label{DiracHamiltonian}
H_\textrm{DF}=\hbar v_F \tau_0\otimes(\boldsymbol{\sigma} \cdot \boldsymbol{p}),
\end{equation}
where $v_F \simeq 10^6 \textrm{ m/s}$ is the Fermi velocity and $\boldsymbol{p} = -i \hbar (\nabla_x,\nabla_y)$ is the momentum operator. $\sigma_{i=x,y,z}$ is the Pauli matrix describing the sublattice degrees of freedom, which is represented by the pseudospin of Dirac fermions. $\tau_{i=0, x, y, z}$ is another Pauli matrix describing the valley ($K$ and $K'$) degrees of freedom; $\tau_0$ is the $2 \times 2$ unit matrix. 


When a graphene sheet has an edge in a certain direction (e.g., armchair or zigzag edge), Dirac fermions are restricted by a boundary condition. In Refs.~\cite{McCann, Akhmerov1}, the boundary condition was derived in the presence of time reversal symmetry and using current conservation at edge. It states that a Dirac-fermion state $\Phi$ at an edge 
%
is proportional to a four-component spinor $\Phi_E$ satisfying the eigenvalue equation of $M \Phi_E = \Phi_E$. The $4 \times 4$ hermitian matrix $M$
is given by
\begin{equation}
M=(\boldsymbol{\nu}\cdot\boldsymbol{\tau})\otimes
[\sigma_z\cos\theta+(\sigma_x\cos\alpha+\sigma_y\sin\alpha)\sin\theta], \label{Matrix_M}
\end{equation}
where $\boldsymbol{\nu}$ is a vector describing intervalley mixing, $\alpha$ is the orientation angle of the edge, and $\theta\in(-\pi/2, \pi/2]$ is determined by physical situations at the edge. 

\begin{figure}[t]
\includegraphics[width=0.47\textwidth]{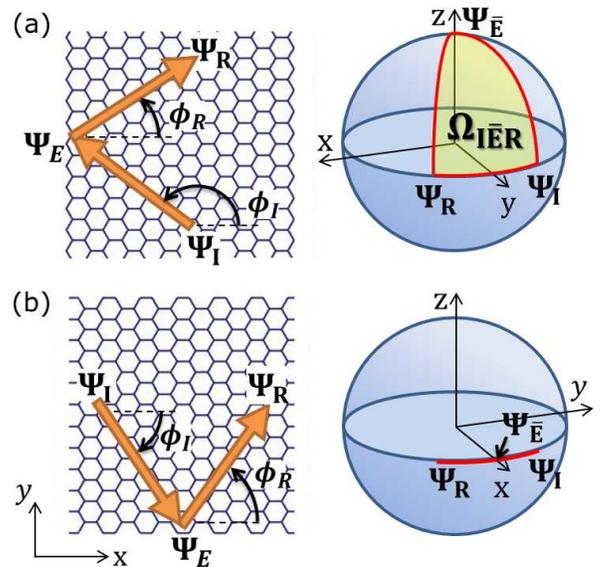}
\caption{ 
Left panel: Reflection of Dirac fermions at (a) zigzag and (b) armchair graphene edge. An incident plane-wave $\Psi_I$ of Dirac fermions in $K$ valley is reflected into the plane wave $\Psi_R$ in $K$ valley at zigzag edge, while reflected into $\Psi'_R$ in $K'$ valley at armchair edge. The reflection is determined by the boundary state $\Psi_E$ at each edge; see Eqs.~\eqref{BC_zigzag} and \eqref{BC_armchair}. Right panel: Geometric phase $\mathcal{P}_{I\widetilde{E}R}$, resulting from the rotation of Dirac-fermion pseudospin in the reflection.
The amount of $\mathcal{P}_{I\widetilde{E}R}$ equals the half of the solid angle on Bloch sphere for pseudospin which is determined by the geodesic lines connecting the pseudospinor of $\Psi_I$, $\Psi_{\widetilde{E}}$, and $\Psi_R$, where $\Psi_{\widetilde{E}}$ is the pseudospinor orthogonal to $\Psi_E$. At zigzag edge, the phase shift is finite, while it vanishes at armchair edge.}
\label{PBphase}
\end{figure}

We present the parameters ($\alpha$, $\boldsymbol{\nu}$, and $\theta$) of the boundary condition for zigzag edge.
When a zigzag nanoribbon is along y-axis [see Fig.~\ref{PBphase}(a)], we choose the orientation angle of $\alpha=\pi/2$. In this case, the intervalley mixing vector is $\boldsymbol{\nu}= t \hat{z}$, where $t= - 1$ ($t= 1$) for the left (right) zigzag ribbon edge. 
$\theta$ depends on physical situations. For a pristine zigzag edge~\cite{Brey,Akhmerov1}, $\theta=0$. For the reconstruction of zigzag-edge structure into energetically favorable pentagon-heptagon pairs~\cite{Koskinen1,Koskinen2,Girit}, a numerical calculation predicts $\theta = 0.15$ ($\theta=-0.15$) at the left (right) zigzag edge of the ribbon~\cite{Ostaay}. In the case of a staggered potential at zigzag edge~\cite{Akhmerov1}, $\theta$ depends on the magnitude and area of the staggered potential.

Under these parameters, the boundary condition of a zigzag edge is obtained from the eigenvector $\Phi_E$ of Eq.~\eqref{Matrix_M}. We present the boundary condition of the left side edge. Since there is no intervalley scattering at a zigzag edge, we consider K and K' valleys independently,
\begin{eqnarray}
\Phi_E & = & A \left( \begin{array}{c} \Psi_E \\ \boldsymbol{0} \\ \end{array} \right) \,\,\,\,\,\, \textrm{for K valley} \nonumber \\ 
\Phi_E & = & B \left( \begin{array}{c} \boldsymbol{0} \\ \Psi'_E \end{array} \right) \,\,\,\,\,\, \textrm{for K' valley,} \label{BC_zigzag}
\end{eqnarray}
where $\boldsymbol{0} \equiv (0, 0)^\dagger$, $\Psi_E^\dagger=(\sin\theta/2, i\cos\theta/2)$, $(\Psi'_E)^\dagger=(\cos\theta/2, -i\sin\theta/2)$, and $A$ and $B$ are coefficients. 
At the right side edge, $\Phi_E$ is found to be orthogonal to that of the left side edge, $\Phi_E^\dagger = ((\Psi'_E)^\dagger, \boldsymbol{0}^\dagger)$ for K valley and $\Phi_E^\dagger = (\boldsymbol{0}^\dagger, \Psi_E^\dagger)$ for K' valley. 

On the other hand, for an armchair edge along $\hat{x}$ axis [see Fig.~\ref{PBphase}(b)], we have $\alpha = 0$, $\boldsymbol{\nu}=(-\cos\gamma, -\sin\gamma, 0)$, and $\theta=\pi/2$ as shown in Refs.~\cite{Brey,Akhmerov1}. The vector $\boldsymbol{\nu}$ lying on $xy$ plane describes the fact that intervalley scattering maximally occurs at armchair edge, and $\gamma$ is the dynamical phase arising from the momentum shift between the two valleys in the scattering. For our purpose of studying the geometric phase by pseudospin rotation, we set aside the dynamical phase by choosing $\gamma=0$, namely, by placing the $y=0$ line at an edge. Besides, in dealing with a metallic armchair ribbon, we may set $\gamma=0$ at the other edge~\cite{Akhmerov1} because the dynamical phase by valley scatterings is $2\pi n$ in that case~\cite{CastroNeto, Brey}.

Under these parameters, we find that the boundary condition at both the upper and lower armchair edges in Fig.~\ref{PBphase}(b) is written as
\begin{equation}
\Phi_E=C\left(
\begin{array}{c}
1\\
0\\
0\\
-1\\	
\end{array}
\right)
+D\left(
\begin{array}{c}
0\\
1\\
-1\\	
0\\
\end{array}
\right).		\label{BC_armchair}
\end{equation}


\section{Pancharatnam-Berry phase at graphene edges}
\label{sec:PBphase}
In this section, we show that Pancharatnam-Berry geometric phase appears in the reflection phase shift of a plane wave of a Dirac fermion at graphene edge.

We consider the situation in Fig.~\ref{PBphase} where a plane wave $\exp(i\vec{k_I}\cdot\vec{r})\Psi_I$
 of Dirac fermions in K valley is incoming, with incident angle $\phi_I$, to an edge and reflected into the plane wave of $\exp(i\vec{k_R}\cdot\vec{r})\Psi_R$ in K valley with angle $\phi_R$ or into $\exp(i\vec{k_{R'}}\cdot\vec{r})\Psi_{R'}$ in K' valley with angle $\phi_{R'}$ ($=\phi_R$); note that $\vec{k_{R'}}$ is identical to $\vec{k_R}$ hence $\phi_{R'} = \phi_R$. The pseudospinors of Dirac fermions are chosen as $\Psi_\mu^\dagger=(\exp(i\phi_\mu/2),\, \exp(-i\phi_\mu/2))^\dagger/\sqrt{2}$, $\phi_\mu=\tan^{-1}[(k_\mu)_y/(k_\mu)_x]$, $\phi_\mu\in(-\pi,\pi]$, and $\mu=I,R,R'$.
For simplicity, we place the origin of the coordinate on a graphene edge (e.g., placing the $x=0$ line on a zigzag edge or the $y=0$ line on an armchair edge; see Fig.~\ref{PBphase}). Then the wavefunction satisfies the continuity equation of 4-component spinors at the edge,
\begin{equation}
\left(
\begin{array}{c}
\Psi_I \\
\boldsymbol{0} \\
\end{array}
\right)+r\left(
\begin{array}{c}
\Psi_R \\
\boldsymbol{0} \\
\end{array}
\right)+r'\left(
\begin{array}{c}
\boldsymbol{0} \\
\Psi_{R'} \\
\end{array}
\right)=\Phi_E,
\label{continuityEQ}
\end{equation}
where $r$ ($r'$) is the reflection amplitude to  the state with pseudospin $\Psi_R$ ($\Psi_{R'}$) in $K$ $(K')$ valley and the boundary condition $\Phi_E$ is given by Eq.~\eqref{BC_zigzag} or \eqref{BC_armchair}, depending on the orientation angle of the edge. 
In the case of an incident plane wave in $K'$ valley, one finds the same equation, except the replacement of $(\Psi_I^\dagger, \boldsymbol{0})^\dagger$ into $(\boldsymbol{0}, (\Psi'_I)^\dagger)^\dagger$. 

In the case of a zigzag edge, we have $r'=0$, since there is no intervalley scattering. Then, Eq.~\eqref{continuityEQ} becomes reduced into an equation for 2-component pseudospinors in K valley,
\begin{equation}
\Psi_I + r\Psi_R = A\Psi_E. \label{zigzag_continuityEQ}
\end{equation}
In this case, we obtain $\textrm{arg} \, r$ in the following steps.
By applying $\Psi_{\widetilde{E}}^{\dagger}$ to both the sides of Eq.~\eqref{zigzag_continuityEQ}, where $\Psi_{\widetilde{E}}^{\dagger}$ is the pseudospinor orthogonal to $\Psi_E^\dagger$, we obtain $r=-(\Psi_{\widetilde{E}}^{\dagger}\Psi_{I})/(\Psi_{\widetilde{E}}^{\dagger}\Psi_{R})$.  Next, we apply the geodesic rule~\cite{Pancharatnam, Berry1} of arg$(\Psi_{a}^{\dagger}\Psi_{b})=i\int_{C}d\vec{s}\cdot\Psi_{\vec{s}}^{\dagger}\nabla\Psi_{\vec{s}}$, where $C:b\rightarrow a$ is the geodesic line from $\Psi_{b}$ to $\Psi_{a}$ on Bloch sphere, to find the relation of $\mathcal{P}_{I\widetilde{E}R}\equiv$arg$[(\Psi_{I}^{\dagger}\Psi_{R})(\Psi_{R}^{\dagger}\Psi_{\widetilde{E}})(\Psi_{\widetilde{E}}^{\dagger}\Psi_{I})]=i\oint_{I\widetilde{E}R}d\vec{s}\cdot\Psi_{\vec{s}}^{\dagger}\nabla\Psi_{\vec{s}}$, where the line integration is done along the geodesic polygon connecting states $\Psi_I, \Psi_{\widetilde{E}}, \Psi_R$ on Bloch sphere. Then, we obtain $|r|=1$ and
\begin{equation}\label{PBphaseEQ}
\textrm{arg } r = \pi - \textrm{arg} \, (\Psi_{I}^{\dagger}\Psi_{R}) + \mathcal{P}_{I\widetilde{E}R}, \,\,\,\,\,\,  \mathcal{P}_{I\widetilde{E}R}=-\frac{\Omega_{I\widetilde{E}R}}{2}.
\end{equation}
$\Omega_p$ is the solid angle covered by the geodesic polygon $p$ (see Fig.~\ref{PBphase}). The first term $\pi$ is the well-known reflection phase by the hard wall boundary condition, and the second term $\textrm{arg} \, (\Psi_{I}^{\dagger}\Psi_{R})$ is a gauge dependent term, which vanishes under the gauge choice of this work. 
The third term $\mathcal{P}_{I\widetilde{E}R}$ is the Pancharatnam-Berry geometric phase and comes from the rotation of the pseudospin in the reflection. This term is gauge invariant, hence, physically meaningful.~\cite{Choi}


For the case of the zigzag edges shown in Fig.~\ref{PBphase}(a), we provide the expression of $\mathcal{P}_{I\widetilde{E}R}$,
\begin{equation}\label{tanPBphase}
\tan\mathcal{P}_{I\widetilde{E}R}=-\frac{t\sin\varphi\cos\theta}{s\cos\varphi-\textrm{sgn}(\varphi)\sin\theta}, 
\end{equation}
where $t=-1$ ($t=1$) for the left (right) zigzag edge, $2\varphi\equiv\phi_R-\phi_I$, 
$\textrm{sgn} \, (\varphi)$ is $1$ $(-1)$ for counterclockwise (clockwise) rotation, and $s=1$ for K valley while $s=-1$ for K' valley. 
For a pristine zigzag edge, we have $\theta=0$ hence $\mathcal{P}_{I\widetilde{E}R}=-ts\,\varphi$. For a zigzag edge with reconstruction and staggered potential, $\theta$ is finite. $\mathcal{P}_{I\widetilde{E}R}$ is finite in general at zigzag edge.

On the other hand, at an armchair edge, an incident plane wave in K valley is reflected into a state in K' valley, and vice versa. Hence, we put $r=0$ in Eq. \eqref{continuityEQ}. By combining Eqs. \eqref{continuityEQ} and~\eqref{BC_armchair}, we find $\Psi_I^\dagger = (C^*, D^*)$ and $(r')^* (\Psi_{R'})^\dagger = (-D^*, -C^*)$, and rearrange them as
\begin{equation}
\Psi_I + r'\Psi_{R'} = (C-D)\left(\begin{array}{c}1\\-1\\ \end{array}\right) \equiv (C-D) \Psi_{E'}. \label{armchair_continuityEQ}
\end{equation}
Here, we introduced $\Psi_{E'}^\dagger = (1, -1)$ to have the same form with Eq.~\eqref{zigzag_continuityEQ}. Then, we obtain
\begin{equation}\label{PBphaseEQ_Armchair}
\textrm{arg } r' = \pi - \textrm{arg} \, (\Psi_{I}^{\dagger}\Psi_{R'}) + \mathcal{P}_{I\widetilde{E'}R'}, \,\,\,\,\,\,  \mathcal{P}_{I\widetilde{E'}R'}=-\frac{\Omega_{I\widetilde{E'}R'}}{2}. \end{equation}
We notice that the Pancharatnam-Berry geometric phase vanishes, $\mathcal{P}_{I\widetilde{E'}R'}=0$ at armchair edges, because $\Psi_I$, $\Psi_{R'}$, and $\Psi_{E'}$ lie on the equator of Bloch sphere; see Fig.~\ref{PBphase}.  
This behavior of armchair edge is very different from the case of zigzag edge.




\section{Geometric phase and band gap of zigzag nanoribbon}
\label{sec:ZGNR}
The contribution of the Pancharatnam-Berry geometric phase to the reflection phase is gauge invariant, and it modifies Bohr-Sommerfeld quantization rule~\cite{Choi}. In this section, we provide an interesting example where the geometric phase affects the electronic band structure and transport of zigzag graphene nanoribbons. We will show that there is connection between the geometric phase and the energy band gap of a zigzag graphene nanoribbon with staggered sublattice potential.

We first discuss the quantization rule for transverse modes in a graphene zigzag nanoribbon. The two edges of the ribbon are located at $x=0$ and $x=W$ in the coordinate of Fig.~\ref{PBphase}. Applying the boundary condition in Eq.~\eqref{zigzag_continuityEQ} to the two edges, we get $a\Psi_I+b\Psi_R=c\Psi_{E_l}$ and $a\exp(-ik_nW)\Psi_I+b\exp(ik_nW)\Psi_R=d\Psi_{E_r}$, where $\Psi_I$ ($\Psi_R$) is the pseudospinor of a plane wave moving along the transverse direction from $x=W$ to $x=0$ (from $x=0$ to $x=W$), $\Psi_{E_{l(r)}}$ is the pseudospinor of the boundary state of the left (right) edge, and $k_n$ is the momentum of the $n$-th transverse mode. 
%
%
These equations are written in an equivalent matrix form
\begin{equation}
\left(\begin{array}{cc}
\,\,\,\,\,\,\,\,\,\,\,\,{\Psi_{\widetilde{E}_l}}^\dagger\Psi_I & \,\,\,\,\,\,{\Psi_{\widetilde{E}_l}}^\dagger\Psi_R \\
e^{-ik_nW}{\Psi_{\widetilde{E}_r}}^\dagger\Psi_I & e^{ik_nW}{\Psi_{\widetilde{E}_r}}^\dagger\Psi_R \\
\end{array}\right)
\left(\begin{array}{c}
a\\ b\end{array}\right)
=\left(\begin{array}{c}
0\\ 0\end{array}\right),
\end{equation}
where $\Psi_{\widetilde{E}_{l(r)}}$ is the pseudospinor orthogonal to $\Psi_{E_{l(r)}}$. The determinant of the matrix is zero, when the matrix equation has a nontrivial solution. Combining this and the geodesic rule~\cite{Pancharatnam, Berry1}, we find the quantization rule of $k_n$,
\begin{equation}\label{quantization}
k_nW = n\pi - \frac{\mathcal{P}_{I\widetilde{E}_l R\widetilde{E}_r}}{2},
\end{equation}
where $\mathcal{P}_{I\widetilde{E}_l R\widetilde{E}_r}=\mathcal{P}_{I\widetilde{E}_l R}+\mathcal{P}_{R\widetilde{E}_r I}$ is the sum of the Pancharatnam-Berry phase accumulated in the reflection processes at the two edges (namely, the total geometric phase resulting from the pseudospin rotation during one period travel of the transverse mode); see Eq.~\eqref{PBphaseEQ}. Notice that $\mathcal{P}_{I\widetilde{E}_l R\widetilde{E}_r} = - \Omega_{I\widetilde{E}_l R\widetilde{E}_r} / 2$ and that $\Omega_{I\widetilde{E}_l R\widetilde{E}_r}$ is the solid angle covered by the geodesic polygon connecting $\Psi_I$, $\Psi_{\widetilde{E}_l}$, $\Psi_R$, and $\Psi_{\widetilde{E}_r}$.

\begin{figure}[t]
\includegraphics[width=0.47\textwidth]{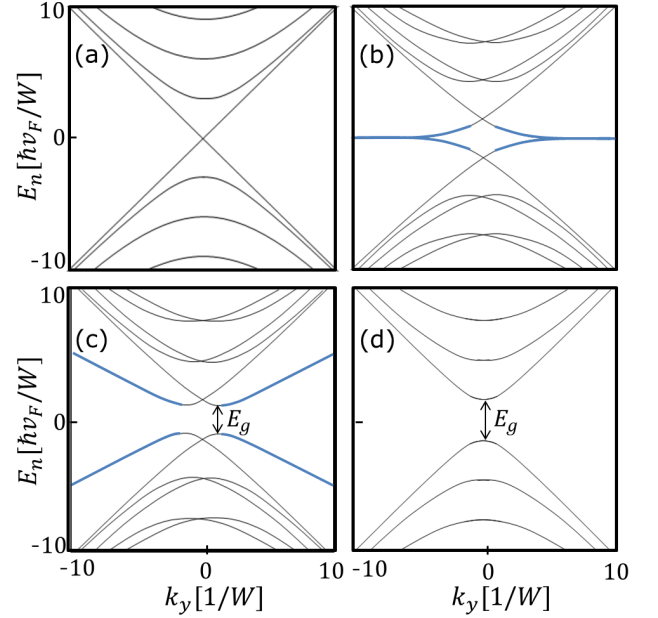}
\caption{
{Electronic band structure of zigzag graphene nanoribbons in the low-energy regime.}
Localized edge states at ribbon edges are marked by thick blue line.
(a) A metallic armchair ribbon. In this case, $\mathcal{P}_{I\widetilde{E}_lR\widetilde{E}_r}=0$ always. (b) A pristine zigzag graphene nanoribbon.
(c-d) Zigzag graphene nanoribbons with the staggered potential of (c) $\theta=\pi/6$ and (d) $\theta=\pi/2$. The transverse modes at band bottoms are pointed by black arrows. $E_g$ denotes the magnitude of the band gap. }
\label{band}
\end{figure}

We point out that in a zigzag nanoribbon, the transverse momentum couples with the longitudinal momentum through the Pancharatnam-Berry phase $\mathcal{P}_{I\widetilde{E}_l R\widetilde{E}_r}$ in the quantization rule of Eq. \eqref{quantization}, as $\mathcal{P}_{I\widetilde{E}_l R\widetilde{E}_r}$ depends on the rotation angle $\varphi$ of pseudospin and $\theta_{l,r}$. Hence the resulting dispersion relation is nonlinear. In contrast, in an armchair nanoribbon, $\mathcal{P}_{I\widetilde{E}_l R\widetilde{E}_r} = 0$ so that the transverse momentum decouples from the longitudinal momentum. This shows that the geometric phase $\mathcal{P}_{I\widetilde{E}_l R\widetilde{E}_r}$ provides the topological view why the electronic structure of a zigzag nanoribbon is very different from that of an armchair nanoribbon, as shown in Figs.~\ref{band}. The difference of the band structures between the nanoribbons (armchair, pristine zigzag, zigzag with edge magnetization) shown in Fig.~\ref{band}(a-d) exhibits the contribution of the geometric phase $\mathcal{P}_{I\widetilde{E}_l R\widetilde{E}_r}$ to the quantization rule.

We apply Eq.~\eqref{quantization} to a zigzag ribbon with pristine edge. In this case, $\mathcal{P}_{I\widetilde{E}_lR\widetilde{E}_r}=2s\varphi$, which is obtained from $\mathcal{P}_{I\widetilde{E}_l R}=\mathcal{P}_{R\widetilde{E}_r I}=s\varphi$; from Eq.~\eqref{tanPBphase},
$\mathcal{P}_{I\widetilde{E}_l R}=s(\phi_R-\phi_I)/2 = s\varphi$ at the left edge (where $t=-1$), while $\mathcal{P}_{R\widetilde{E}_r I}=-s(\phi_I-\phi_R)/2=s(\phi_R-\phi_I)/2=s\varphi$ at the right edge (where $t=1$). 
 Then, the quantization rule in Eq.~\eqref{quantization} is given by $k_y=-s\tan(\varphi)=sk_n/\tan(k_nW)$, where $k_n$ and $k_y$ are the transverse and longitudinal momentum wave vectors, respectively. This reproduces the quantization rule found in Ref.\cite{Brey}.

We also apply Eq.~\eqref{quantization} to a zigzag nanoribbon with staggered sublattice potential~\cite{Akhmerov1} at edges, which describes edge magnetization and results in band gap opening~\cite{Yamashiro, Son}. Below, for a symmetric magnetization case of $\theta_r = \theta_l$ at the ribbon edges, we further derive the quantization rule of Eq.~\eqref{quantization}. 
Using Eq. \eqref{tanPBphase} and applying $\theta = \theta_r = \theta_l$, we get $\tan\mathcal{P}_{I\widetilde{E}_l R\widetilde{E}_r}=\tan(\mathcal{P}_{I\widetilde{E}_l R}+\mathcal{P}_{R\widetilde{E}_r I})=2(s\tan\varphi/\cos\theta)/[1-(s\tan\varphi/\cos\theta)^2]$. Comparing this with $\tan\mathcal{P}_{I\widetilde{E}_l R\widetilde{E}_r}=2(\tan\mathcal{P}_{I\widetilde{E}_l R\widetilde{E}_r}/2)/[1-(\tan\mathcal{P}_{I\widetilde{E}_l R\widetilde{E}_r}/2)^2]$, we obtain the explicit expression of the Pancharatnam-Berry phase, $\tan\mathcal{P}_{I\widetilde{E}_l R\widetilde{E}_r}/2=s\tan\varphi/\cos\theta$. Then, applying $\tan\varphi=-k_n/k_y$, we obtain the explicit form of Eq.~\eqref{quantization} for a symmetric zigzag nanoribbon with staggered sublattice potential,
\begin{equation}\label{tan_quantization}
k_y = \frac{s}{\cos\theta}\frac{k_n}{\tan k_nW}.
\end{equation}
We point out that Eq.~\eqref{tan_quantization} determines the quantization rule and the band structure of \textit{extended} modes in the transverse direction. Applying the analytic continuation of $k_n \rightarrow ik_n$ into Eq. \eqref{tan_quantization}, we also obtain the band structure of the localized states at the ribbon edges in the transverse direction of the ribbon; the localized states have non-negligible probability near only one of the two edges of the ribbon. The resulting band structures are shown in Figs.~\ref{band}(c) and (d).

The magnitude of the band gap is determined by the lowest positive energy (the band bottom) of the $n=0$ band [see Figs.~\ref{band}(c)-(d)]. The transverse momentum $k_{y,g}$ at the band bottom satisfies $\partial E(k_{n=0}, k_{y}) / \partial k_y |_{k_{y,g}} = 0$, namely, $k_{y,g} + k_{n=0} d k_{n=0} / d k_y |_{k_{y,g}} = 0$, where $E(k_n, k_y) = \hbar v_F \sqrt{k_n^2 + k_y^2}$. After some algebra, and utilizing Eq.~\eqref{tan_quantization}, this condition is rewritten as a transcendental equation of $k_{n=0} W = (1+\cos^2\theta\tan^2k_{n=0}W)\sin k_{n=0}W \cos k_{n=0}W$. From this equation, Eq.~\eqref{tan_quantization}, and Eq.~\eqref{quantization}, we obtain the transverse momentum $k_{y,g}$, and find the connection between the band gap $E_g$ and the Pancharatnam-Berry phase $\mathcal{P}_{I\widetilde{E}_lR\widetilde{E}_r} (k_{n=0}, k_{y,g})$ occurring in the reflections of the plane wave with momentum $(k_{n=0}, k_{y,g})$ at the two edges,
\begin{equation}\label{bandgap}
E_g=\frac{\hbar v_F}{W}|\mathcal{P}_{I\widetilde{E}_lR\widetilde{E}_r}|\sqrt{1+\frac{\sec^2\theta}{\tan^2(\mathcal{P}_{I\widetilde{E}_lR\widetilde{E}_r})/2}}.
\end{equation}
This shows that the energy band gap of a zigzag graphene nanoribbon connects with the geometric phase. 

\begin{figure}[t]
\includegraphics[width=0.47\textwidth]{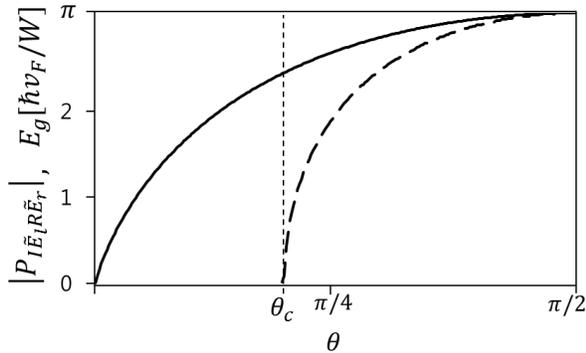}
\caption{
{The band gap (solid line) and the Pancharatnam-Berry phase (dashed) of a zigzag nanoribbon with staggered potential, as a function of $\theta$.}
In the case 
of $\theta=\pi/2$, $E_g=\frac{\hbar v_F}{W}|\mathcal{P}_{I\widetilde{E}_lR\widetilde{E}_r}|$ and $\mathcal|{P}_{I\widetilde{E}_l R\widetilde{E}_r}|=\pi$.
}
\label{Fig_bandgap}
\end{figure}

In Fig. ~\ref{Fig_bandgap}, we draw the magnitude of band gap and the Pancharatnam-Berry phase as a function of $\theta$. We point out that the Panchartanm-Berry geometric phase is not applicable in the regime of $\theta < \theta_c=\cos^{-1}\sqrt{2/3}$ where the band gap is determined by the edge states localized at the edges of the zigzag nanoribbon (i.e., the band bottom state is a localized state), since the geometric phase is defined for the extended states propagating in the transverse direction of the ribbon.

\section{Veselago lens in graphene nanoribbon}
\label{sec:Veselago}

Veselago lens is an optical device of negative refractive index that focuses light.~\cite{Veselago} Its electronic analogue~\cite{Cheianov} can be realized in a graphene $p$-$n$ junction; a graphene $p$-$n$ junction and the type of charge carriers are controllable by gate voltages or doping.~\cite{Novoselov2,Ohta}
Negative refraction of electron flow occurs at the junction interface. When the carrier density is identical between the $p$ and $n$ regions, the junction exhibits perfect focusing where electron flow converges at the focal point of the electron injection by a source tip.
When the carrier density is unequal between the $n$ and $p$ regions, on the other hand, the perfect focusing does not occur, but electron interference leads to caustics pattern near the junction.~\cite{Cheianov} 
In this section, we will numerically calculate caustics pattern in a Veselago lens formed in a graphene nanoribbon, and show that the interference pattern depends on the edge structure of the ribbon, resulting from the reflection phase shift at the edges, namely from the Pantanratnam-Berry geometric phase. This provides a direct way of detecting the geometric phase at graphene edge.

This section has two subsections. In subsection A, we introduce the model system and the calculation method based on Green functions. In subsection B, we discuss the caustics pattern for various situations of a zigzag nanoribbon. We compare the result with an armchair nanoribbon where the geometric phase is absent. 

\subsection{Calculation method}

To compute the caustics pattern, we use a tight-binding lattice Hamiltonian; the result of the tight-binding calculation supports the prediction of the Pancharatnam-Berry geometric phase based on the continuum Hamiltonian for Dirac fermions. The tight-binding Hamiltonian $H = H_0 + H_T$, describing a $p$-$n$ junction in a graphene nanoribbon and a source tip on the $n$ region, is written as
\begin{eqnarray}
H_0 & = & -t \sum_{\langle i, j \rangle} c^{\dagger}_i c_j + \sum_{i}\epsilon_i c^{\dagger}_i c_i + \sum_{k}\epsilon_k d^{\dagger}_k d_k \nonumber \\
H_T & = & \sum_{l, k}(t_s c^{\dagger}_l d_k + h.c.). \label{veselagoHamiltonian} \end{eqnarray}
Here, $\langle i, j\rangle$ means the summation over pairs of nearest neighboring lattice sites $i$ and $j$ in the two-dimensional honeycomb lattice of graphene. $t$ ($=3.090$ $e$V) is the nearest-neighbor hopping energy, and $c^{\dagger}_i$ ($c_i$) creates (annihilates) an electron in the $\pi$ orbital of site $i$ with on-site energy $\epsilon_i$. $\epsilon_i$ is set by $\epsilon_i = eV_e$ ($\epsilon_i=eV_h$) for lattice sites in the $n$ ($p$) region. The source tip is described by a free-electron model. $d^{\dagger}_k$ $(d_k)$ creates (annihilates) an electron with momentum $k$ and energy $\epsilon_k$ in the tip. $H_T$ describes electron tunneling, with amplitude $t_s$, between the tip and the lattice sites $l$ of the nanoribbon. The summation over $l$ accounts the resolution of the tip. We place the tip center on a position of the $n$ region, choosing the tunneling sites $l$ as the six nearest neighboring sites from the tip center (that form a smallest hexagon cell in the honeycomb lattice),~\cite{Xing} and study the resulting caustics pattern in the $p$ region. And, we omit the spin degree of freedom, as it provides only two-fold degeneracy. 

To study the caustics pattern, we consider the change of the local electron density at the sites of the $p$ region by the bias voltage $\delta V$ applied to the tip at zero temperature. It is numerically calculated in the linear response regime, based on the Hamiltonian~\eqref{veselagoHamiltonian} and the recursive Green function method~\cite{Xing, Sim}. The change of the local electron density $\delta\rho_i$ at site $i$ is obtained~\cite{Xing} as
\begin{eqnarray}
\delta\rho_i / \delta V =\frac{1}{2\pi}[\bold{G}^r(E_F)\bold{\Gamma}_s(E_F) \bold{G}^a(E_F)]_{ii},
\end{eqnarray}
where $E_F$ is the Fermi energy.
The quantities with boldface are matrices, represented in the lattice-site basis. $\bold{G}^r(E)$ is the retarded Green's function obtained from $H_0$, while $\bold{G}^a(E) = [\bold{G}^r(E)]^\dagger$ is the advanced Green function. $\bold{\Gamma}_s$ is the linewidth function of the source tip, and it is given by $\bold{\Gamma}_s=2\pi t^2_s\rho_s(E_F)\bold{I}_s$ in the wide-band approximation~\cite{Lopez}. Here, $\rho_s$ is the density of states of the tip,
$\bold{I}_s$ is the matrix describing the tunneling between the tip and the sites $l$ of the nanoribbon, $[\bold{I}_s]_{mn}=\Sigma_l \delta_{ml}\delta_{nl}$, and $\delta_{ab}$ is the Kronecker delta function. Note that to study a nanoribbon with infinite length, $\bold{G}^r(E)$ is computed by using the recursive Green function method~\cite{Xing,Sim}; in this method, the nanoribbon is divided into a left semi-infinite part, a right semi-infinite part, and a middle part containing the $p$-$n$ junction, and the Green functions of the two semi-infinite parts are computed separately and affect the Green function of the middle part as a self energy. 

In our calculation, we choose $E_F=0$, the width $W$ of the nanoribbon as $W\simeq150\textrm{nm}$, and the gate voltages as $eV_e=-0.083t$ and $eV_h=0.10t$; under these gate voltages, the refractive index of the junction is $N = V_h/V_e=-1.2$. Note that we obtained the average of $\delta\rho_i / \delta V$ over the nearest neighboring six lattice sites of site $i$, considering the resolution limit of a local detector of charge density change.~\cite{Xing} The average washes out rapid oscillations of $\delta\rho_i / \delta V$ that vary over the lattice constant.
 


\begin{figure}[t]
\includegraphics[width=0.46\textwidth]{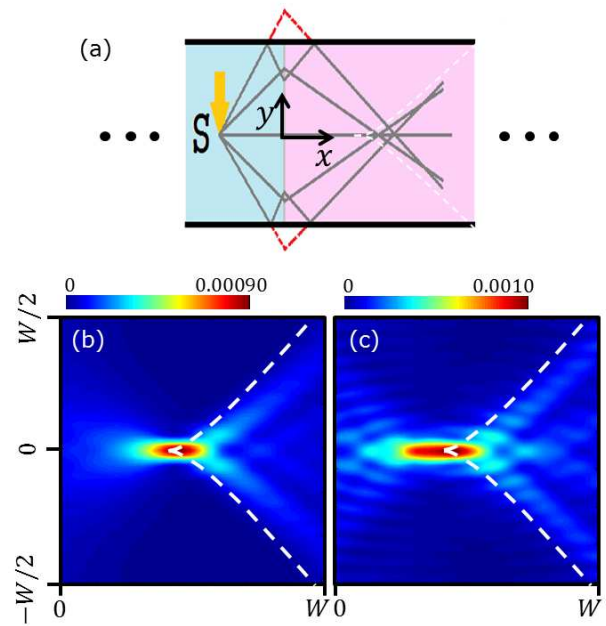}
\caption{(Color online)
(a) Veselago lens in a graphene nanoribbon with a source tip. The $p$-$n$ junction for the lens lies at $x=0$. The left side of the junction is $n$-doped, while the right is $p$-doped. The tip is placed on a center position (marked by the yellow arrow) in the transverse direction in the $n$ region. We choose the tip position at $S=(-a, 0)$, $a=48\textrm{nm}$; see the yellow arrow. The gray lines show classical trajectories of the electron injected from the tip in the case of refractive index $N<-1$. The trajectories converge around the focal point, while some trajectories are reflected twice at edges before the converge. The edges of the ribbon are placed at $y = \pm W / 2$, where the ribbon width $W$ is chosen as $W = 150$ nm. (b,c) Caustics pattern in the $p$ region of $x>0$ in the cases of (b) a metallic armchair nanoribbon and (c) a zigzag nanoribbon with pristine edges. The pattern shows the change of the local electron density $\delta \rho_i$ by electron injection from the tip. Dashed lines show the caustic curves, and the position of focal point is $f=(|na|,0)$.
}
\label{ldosCenter}
\end{figure}

\subsection{Caustics pattern}

We first discuss the caustics pattern of the case where the source tip is placed exactly on the center position of the nanoribbon in the transverse direction. Figure~\ref{ldosCenter}(a) shows the setup and the caustic curves; for the formula for the caustic curves, see Ref. ~\cite{Cheianov}.

In the case of an armchair nanoribbon, the Pancharatnam-Berry phase at the edges vanishes because of intervalley mixing, hence, does not have an impact on the caustics pattern. Indeed, as shown in Fig.~\ref{ldosCenter}(b), the caustics pattern has the same form as that of a Veselago lens formed in an infinite-size graphene, although some electron trajectories are reflected twice at one of the edges in the nanoribbon case.


On the other hand, in the case of a zigzag nanoribbon, the caustics pattern is affected by the Pancharatnam-Berry phase. In Fig.~\ref{ldosCenter}(c), the caustics pattern is drawn for a zigzag nanoribbon with pristine edges. In this case, the total reflection phase along a trajectory undergoing edge reflection twice is obtained from the geometric phase in Eq.~\eqref{tanPBphase} as  $2\pi - s\varphi_e-s\varphi_h$ ($\equiv - s(\varphi_e+\varphi_h)$ in mod $2\pi$), where $\varphi_e$ ($\varphi_h$) is the angle of rotation of pseudospin in the edge reflection of $n$ ($p$) region.
This modifies interference fringes. As a result, the caustics pattern is different from that of the armchair case.

\begin{figure}[t]
\includegraphics[width=0.46\textwidth]{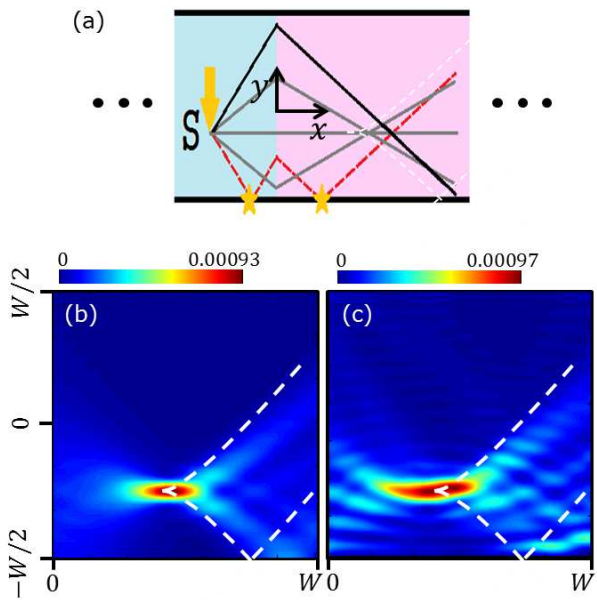}
\caption{(Color online)
Veselago lens in a graphene nanoribbon with a source tip at an eccentric point. The same as Fig.~\ref{ldosCenter}, except that the tip position is located at $S=(-a, d)$, $a=48\textrm{nm}$ and $d=-W/4$. As a result, the focal point $f=(|na|,d)$ is also different from Fig.~\ref{ldosCenter}, and more importantly, the trajectories become asymmetric such that some trajectories propagating into the lower edge undergo edge reflection twice (see the two asterisks) before the converge to the focal point, while their symmetry partners with respect to the $y = d$ line, propagating into the upper edge, do not meet the edge. The case of a metallic armchair nanoribbon is in (b), while the case of a zigzag nanoribbon with pristine edges is in (c).
}
\label{ldosEccentric}
\end{figure}

We next discuss the caustics pattern of the case where the source tip is placed on an eccentric position in the transverse direction; see Fig.~\ref{ldosEccentric}.  In the armchair case, the caustics pattern within the caustics curves (see the white dashed lines in Fig.~\ref{ldosEccentric}) is just shifted along the change of the tip position without any deformation. In contrast, the pattern is deformed into an asymmetric shape in the zigzag case. Hence, the difference of the caustics pattern between the armchair and zigzag cases is more pronounced, so it may be useful for detecting the Pancharatnam-Berry phase.

The reason of the asymmetric pattern in the zigzag case is as follows. When the tip position moves from the center line of $y=0$ into $y=d$, electron trajectories injected from the tip become asymmetric such that some trajectories propagating into the lower edge undergo edge reflection twice before the converge to the focal point, while their symmetry partners with respect to the $y = d$ line, moving into the upper edge, do not meet the edge. Then, the trajectories downward with the edge reflections have phase shift by the Pancharatnam-Berry geometric phase, while those upward have no phase shift. This makes the caustics pattern asymmetric. The asymmetric pattern is a direct signature of the geometric phase. 



We note that under the edge reconstruction or the staggered potential, the caustics pattern of a zigzag nanoribbon remains asymmetric (hence, still in sharp contrast to an armchair nanoribbon) when the tip is on an eccentric point; we obtained numerical results for this case, but do not present them in this paper. Therefore, the caustics pattern provides a tool for detecting the geometric phase.



\section{Summary}
\label{sec:Summary}

In summary, we predict that Pancharatnam-Berry geometric phase appears in the reflection phase of Dirac fermions at graphene edge. The geometric phase is originated from the rotation of pseudospin in the edge reflection. It depends on edge chirality, the injection angle of Dirac fermions into edge, and the detailed edge situation such as edge reconstruction and staggered potential. It is finite in general at zigzag edge, while it always vanishes at armchair edge. As physical manifestation of the geometric phase, we discuss the electronic structure and the energy band gap of graphene nanoribbon. The quantization rule of the transverse mode of the nanoribbon is modified by the geometric phase. And, we also discuss the caustics pattern in a Veselago lens formed in a graphene nanoribbon. The shape of the pattern is different between an armchair nanoribbon and a zigzag nanoribbon, and provides a direct signature of the geometric phase.
Our finding reveals the topological aspect of scattering of Dirac fermions at graphene edge. It can be detected in a resonance or interference setup, as we have discussed the quantization rule and the caustics pattern in a nanoribbon.



\section*{ACKNOWLEDGMENTS}

This work was supported by Korea NRF (Grant No. 2013R1A2A2A01007327).



\end{document}